\documentclass[%
 reprint,
superscriptaddress,
 amsmath,amssymb,
 aps,
]{revtex4-1}

\usepackage{graphicx}% Include figure files
\newcommand{\degree}{$^{\circ}$}

\begin{document}

\title{Multi-metallic conduction cooled superconducting radio-frequency cavity with high thermal stability}% Force line breaks with \\

\author{G. Ciovati}
 \email{gciovati@jlab.org}
 \affiliation{Thomas Jefferson National Accelerator Facility, Newport News, VA 23606, USA}
 \affiliation{Center for Accelerator Science, Department of Physics, Old Dominion University, \\Norfolk, Virginia 23529, USA}
\author{G. Cheng}
 \affiliation{Thomas Jefferson National Accelerator Facility, Newport News, VA 23606, USA}
\author{U. Pudasaini}
\affiliation{The College of William \& Mary, Williamsburg, VA 23185, USA}
\author{R. Rimmer}
\affiliation{Thomas Jefferson National Accelerator Facility, Newport News, VA 23606, USA}
% Authors' institution and/or address\\ This line break forced with \textbackslash\textbackslash

\date{\today}% It is always \today, today,

\begin{abstract}

Superconducting radio-frequency cavities are commonly used in modern particle accelerators for applied and fundamental research. Such cavities are typically made of high-purity, bulk Nb and are cooled by a liquid helium bath at a temperature of $\sim 2$~K. The size, cost and complexity of operating a particle accelerator with a liquid helium refrigerator makes the current cavity technology not favorable for use in industrial-type accelerators. We developed a multi-metallic 1.495~GHz elliptical cavity conductively cooled by a cryocooler. The cavity has a $\sim 2$~$\mu$m thick layer of Nb$_3$Sn on the inner surface, exposed to the rf field, deposited on a $\sim 3$~mm thick bulk Nb shell and a bulk Cu shell, of thickness $\geqslant 5$~mm deposited on the outer surface by electroplating. A bolt-on Cu plate 1.27~cm thick was used to thermally connect the cavity equator to the second stage of a Gifford-McMahon cryocooler with a nominal capacity of 2~W at 4.2~K.
The cavity was tested initially in liquid helium at 4.3~K and reached a peak surface magnetic field of $\sim 36$~mT with a quality factor of $2\times 10^9$. The cavity cooled by the crycooler achieved a peak surface magnetic field of $\sim 29$~mT, equivalent to an accelerating gradient of 6.5~MV/m, and it was able to operate in continuous-wave with as high as 5~W dissipation in the cavity for 1~h without any thermal breakdown. This result represents a paradigm shift in the technology of superconducting accelerator cavities.
\end{abstract}

\maketitle

%\tableofcontents

Superconducting radio-frequency (SRF) cavities made of high-purity (residual resistivity ratio, $RRR$, $> 250$) bulk Nb are one of the building blocks of modern particle accelerator facilities for applied and fundamental research throughout the world \cite{Padamsee}. Such cavities have different geometries, depending on the speed and type of particle they are designed to accelerate, the operating frequency is in the gigahertz range and they are surrounded by vessels containing liquid He (LHe) which cools and maintains the cavity surface at $\sim 2$~K during operation in a so-called cryomodule \cite{RAST_hibeta, RAST_lowbeta}. The size, cost and complexity of a sub-cooled liquid He cryoplant has limited a more widespread application of SRF technology so far. To the authors' knowledge, out of the estimated $\sim 30,000$ industrial particle accelerators worldwide, the only one using the SRF technology is a 9~MeV electron linac for medical isotope production which uses a commercial liquid He refrigerator with a capacity of 100~W at 4.3~K \cite{Niowave}.

The first application of cryocoolers for SRF cryomodules was done in two cryomodules for the Japan Atomic Energy Research Institute Free Electron Laser in 1993, in which they used Gifford-McMahon (GM) cryocoolers to cool two heat shields to 80~K and 40~K, respectively, and to cool down and recondense the boiled-off liquid in the helium tank surrounding a 499.8 MHz cavity \cite{JAERI1, JAERI2}. Cryocoolers are reliable, compact, closed-cycle refrigerators which are less expensive and easier to operate than LHe ones. An example of a cryocooler application is cooling of superconducting magnets in magnetic resonance imaging machines at hospitals. The power capacity of cryocoolers has been increasing in the last few years and models with a capacity of 2~W at 4.2~K are now available.

Recent progress in the development of thin-film Nb$_3$Sn SRF cavities has resulted in moderate accelerating gradients ($E_{acc} \sim 10-15$~MV/m) but a much higher quality factor ($Q_{0} \sim 10^{10}$) compared to bulk Nb in elliptical cavities cooled in liquid He at 4.3~K \cite{Posen_2017}. Such improvements in both cryocooler and Nb$_3$Sn SRF cavities may enable the design of compact, low-energy ($1 - 25$~MeV) electron accelerators for industrial and medical applications or for compact light sources \cite{Kephart, CLS_Report}. An example of such industrial applications is the environmental remediation of flue gases and/or wastewater \cite{EnvAcc}. Recent work on the development of SRF cavities conduction cooled by a cryocooler has resulted in a 650~MHz single-cell elliptical cavity made of bulk Nb operating up to an accelerating gradient of 1.5~MV/m \cite{Dhuley1}, corresponding to a peak surface magnetic field, $B_p$, of 5.5~mT \cite{DhuleyPC}. %and a dissipated power, $P_{loss}$ of $\sim 4$~W.

Here we describe the preparation and test results of a single-cell elliptical cavity conduction cooled by a two stage commercial GM cryocooler. Our approach was to deposit a thick, high-purity Cu layer on the outer cavity surface and to minimize the number of joints between the cavity and the 4~K stage of the cryocooler, to maximize the thermal stability of the cavity against quenching of the superconducting state.

The single-cell cavity used for this study was made of large-grain Nb ($RRR \sim 280$) from CBMM, Brazil. The cell shape is that of the end-cell of a High-Gradient cavity (geometry factor, $G = 269~\Omega$, $R/Q = 100.3~\Omega$, ratio of peak surface electric field over the accelerating field, $E_p/E_{acc} = 1.77$,  ratio of peak surface magnetic field over the accelerating field $B_p/E_{acc} = 4.47$~mT/(MV/m)), proposed for the 12~GeV Upgrade of the CEBAF accelerator at Jefferson Lab  \cite{HG}. The resonant frequency of the TM$_{010}$ accelerating mode is 1.495~GHz. The cavity wall thickness is $\sim2.9$~mm and the end flanges are made of pure Nb. The cavity fabrication used standard techniques of SRF technology such as deep-drawing, milling and electron-beam welding of cavity parts.

The coating of the inner surface of the cavity with Nb$_3$Sn was done together with another single-cell cavity, stacked vertically inside a high-temperature vacuum furnace. The cavity used for this study was at the bottom position. A crucible with 6~g of Sn (99.999\% purity from Sigma Aldrich) and 3 g of SnCl$_2$ (99.99\% purity from Sigma Aldrich), packaged inside two pieces of Nb foils, was placed at the bottom flange of the bottom cavity. The top flange of the top cavity was closed with a Nb cover. The setup was assembled inside an ISO 4 clean room and then installed onto the furnace insert \cite{Nb3Sn_furnace}. Once the pressure reached $2.7 \times 10^{-3}$~Pa, the furnace was heated by ramping up the temperature at a rate of 6~\degree C/min until it reached $\sim500$ \degree C. This temperature was then kept constant for one hour and subsequently ramped up at a rate of 12~\degree C/min up to the coating temperature of $\sim 1200$~\degree C. The temperature was monitored with sheathed type C thermocouples attached to the cavities at different locations. After maintaining the coating temperature for 3 h, heating ceased, and the furnace was allowed to cooldown gradually. When the furnace temperature reached below 45~\degree C, the insert was backfilled to 101.3~kPa with nitrogen, and the coated cavities were removed from the deposition system.
The cavity was then degreased in an ultrasonic tank, high-pressure rinsed with ultrapure water, assembled with stainless steel flanges with pump-out port and rf feedthroughs and sealed to the cavity with In wire. The cavity was evacuated on a vertical test stand to a pressure of $\sim1\times10^{-6}$~Pa before inserting in a vertical cryostat in the Vertical Test Area (VTA) at Jefferson Lab. Cryogenic flux-gate magnetometer (FGM) probes (Mag F, Bartington Instruments) and calibrated Cernox (CX-1010-SD, Lakeshore Cryotronics) resistance-temperature devices (RTDs) were attached to the cavity to monitor the temperature gradient along the cavity and the local magnetic flux during cooldown close to the critical temperature of Nb$_3$Sn, $T_c \sim 18$~K.
%The rf performance of the cavity in liquid He after the first coating with Nb$_3$Sn was limited by anomalous heating at 4.3~K starting at $B_p \sim 40$~mT and by thermal quench at $B_p \sim 66$~mT at 2.0 K. In both cases, the quality factor at the highest rf field was $\sim 6 \times 10^9$.
%The Nb$_3$Sn film was removed by chemical etching and the coating was repeated in the same way as described above, except that the cavity position on the furnace insert was switched. A visual inspection of the coated cavity showed a uniform coating, but a few shiny spots were present on the surface. 
The rf performance of the Nb$_3$Sn cavity in liquid He at 4.3~K was limited by anomalous heating at 4.3~K starting at $B_p \sim 36$~mT and by thermal quench at $B_p \sim 54$~mT at 2.0 K.

Oxygen-free high-conductivity (OFHC) copper is one of the metals which has a higher thermal conductivity than high-purity Nb below 10~K and it has been used as a substrate for the deposition of Nb thin-films on the inner surface of cavities for particle accelerators \cite{LEP}. The higher thermal conductivity allows for a better thermal stabilization of the cavity, even when cooling with LHe, particularly against the presence of defects in the superconducting thin-film. However, at present there is no technique which allows depositing a thin film of Nb$_3$Sn directly onto copper with similar performance as achieved by forming the Nb$_3$Sn layer onto bulk Nb by vapor diffusion.

Nb/Cu bi-metallic samples were produced by electroplating Cu directly onto a Nb and thermal conductivity measurements showed that values of $\sim 1$~kW/(m$\cdot$K) could be achieved at 4.3~K, compared to $\sim 75$~W/(m$\cdot$K) obtained on Nb only \cite{Gigi_SRF19}. However, achieving such high thermal conductivity depended on obtaining a good adhesion of the Cu on the Nb, which we were not able to achieve reliably and consistently by electroplating Cu directly onto the Nb. A recent collaboration between Jefferson Lab, Euclid Techlabs and Concurrent Technologies Corporation (CTC) produced Nb/Cu samples obtained by cold-spraying Cu onto Nb with excellent bonding: a pressure of $\sim 40$~MPa was required to detach the Cu from the Nb in a pull-adhesion test \cite{Euclid}. However, the thermal conductivity of the ``as deposited`` cold-sprayed copper was not as high as that obtained by electroplating. While R\&D efforts are ongoing towards increasing the thermal conductivity of cold-sprayed copper, we pursued cold-spraying as a method to grow a thin seed layer onto the Nb and then electroplate the copper to full thickness on such layer. An iterative finite element thermal analysis of the cavity-cryocooler system was carried out with ANSYS~\cite{ansys} estimating that a minimum Cu-layer thickness of 5~mm would be necessary in order to achieve a $B_p$-value of $\sim 36$~mT. The analysis included the heat capacity map of the cryocooler, the temperature- and field-dependent surface resistance of the Nb$_3$Sn cavity and a contact thermal conductance of 0.7~W/K \cite{Dillon_2017}.

\begin{figure}[htb]
\includegraphics*[width=85mm]{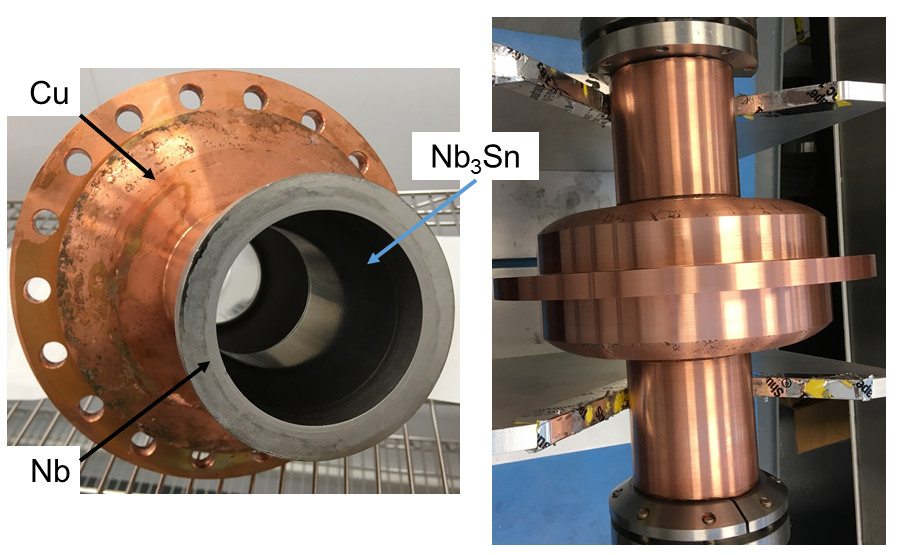}
\caption{\label{fig:cav} Pictures of the multi-metallic SRF cavity. The length of the cavity from flange to flange is 31.8~cm and the electroplated Cu ring at the equator is 25.4 cm in diameter and 1.27~cm thick.}
\end{figure}

A copper layer $\sim 76$~$\mu$m thick was deposited on the cavity outer surface by cold-spray at CTC, in Johnstown, PA. Copper powder of 99.9~\% purity and $\sim 40$~$\mu$m size was used along with He as gas carrier. Oxygen-free copper was then electroplated onto the cold-sprayed layer at AJ Tuck Co., in Brookfield, CT. The electroplating was done in several steps to control the uniformity of the thickness and finally to grow a ring $\sim 25$~cm in diameter, $\sim 1.3$~cm thick at the cavity equator. The cavity was finally machined at Jefferson Lab to remove excess Cu and to make eighteen holes evenly spaced along the Cu ring at the cavity equator. Gore-Tex gaskets were used to seal the cavity ends during both cold-spraying and electroplating, however it was found that some of the copper sulfate plating solution had leaked inside the cavity. The cavity was filled with nitric acid at room temperature for 1~h to dissolve any possible CuSO$_4$ residue. Figure~\ref{fig:cav} shows a picture of the completed multi-metallic cavity.
The cavity was prepared for the cryogenic rf test as decribed earlier and the performance measured in liquid He was limited by ``Q-switches`` at $B_p \sim 35$~mT at 4.3~K, whereas it quenched at $B_p \sim 52$~mT at 2.0 K. However, the quality factor degraded more rapidly with increasing field above $\sim 14$~mT, compared to the test prior to Cu-coating.

\begin{figure}[htb]
\includegraphics*[width=86mm]{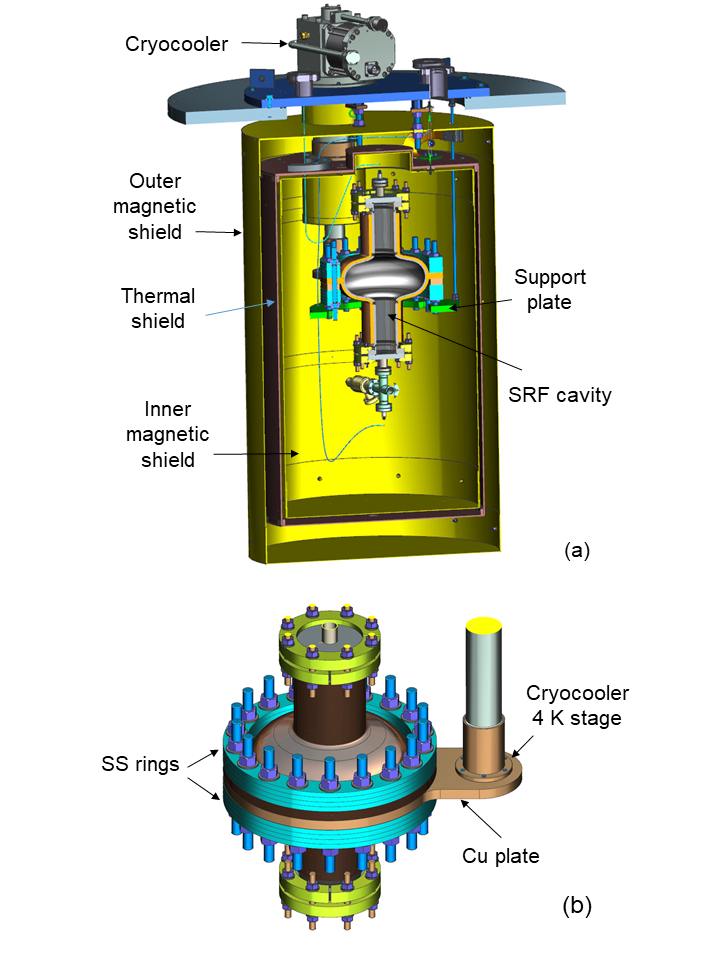}
\caption{\label{fig:stand} Cross-section of the 3D model of the cavity-cryocooler the test stand (a) and detail of the cavity connection to the cryocooler 4~K stage (b). The test stand has a 61~cm diameter top plate and it is 93 cm in height.}
\end{figure}

A vertical test stand was designed and built to allow testing the cavity with a cryocooler. The GM cryocooler (RDE-418D4, Sumitomo) is bolted to the test stand top plate. The cavity is kept under static vacuum and sits on a G10 plate held by two stainless steel threaded rods attached to the top plate. A plate $\sim 1.27$~cm thick machined from OFHC copper is bolted to the crycoooler 4~K stage on one side and to the cavity equator ring on the other side. The contact surfaces were cleaned with Brasso metal polish and wiped with acetone and isopropanol. Apiezon N thermal grease was spreaded on the contact surfaces. The bolts connecting the Cu plate to the cryocooler were torqued to 3~N$\cdot$m, as recommended by the cryocooler manufacturer. The cavity Cu ring and the Cu plate were sandwiched between four 304 stainless steel rings, each 0.64~cm thick, on each side and pressed together with 1.27 cm diameter 316 stainless steel threaded rods and silicon-bronze nuts, torqued to 115~N$\cdot$m. Such combination of number of rings and torque value allowed achieving a uniform pressure along the ring, estimated to be $\sim 46$~MPa. A high, uniform pressure allows minimizing the thermal resistance of the joint. Figure~\ref{fig:stand} shows a 3D rendering of the test stand and of the cavity connected to the cryocooler. All of the stainless steel hardware had been properly degaussed prior to installation onto the cavity.

The cavity and cryocooler 4~K stage were wrapped with ten layers of multi-layer insulation (MLI) and they are inside an inner magnetic shield. Such inner shield is inside a copper cylinder thermal shield attached to the cryocooler 50~K stage and wrapped with ten layers of MLI. Finally, an outer magnetic shield surrounds the thermal shield. Two rf cables connect the input and pick-up antennae mounted on the cavity to rf feedthroughs on the top plate. Sixteen calibrated Cernox RTDs were distributed on the cavity, Cu plate and the thermal shield. Three cryogenic flux-gate magnetometer probes were placed at locations on the cavity ring with different orientations. 

\begin{figure}[htb]
\includegraphics*[width=86mm]{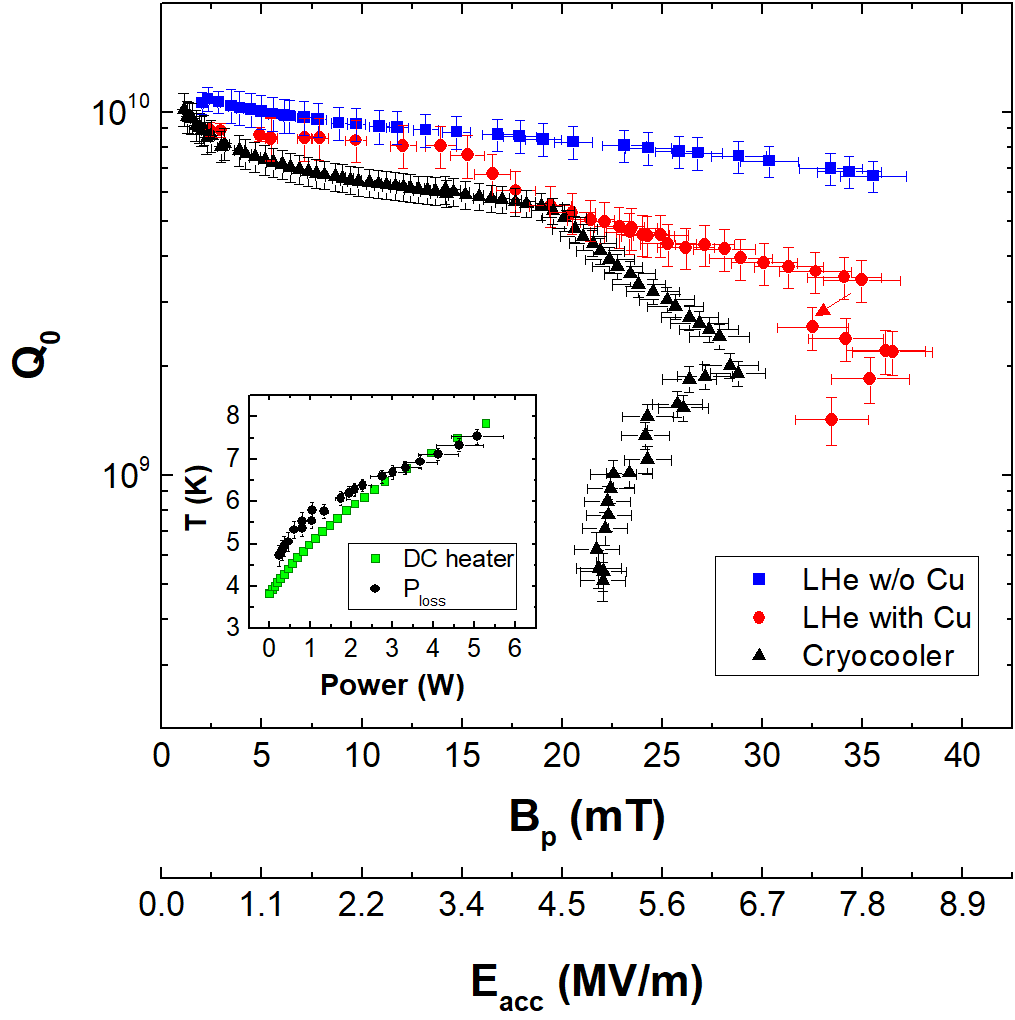}
\caption{\label{fig:rf} Quality factor of the SRF cavity as a function of the peak surface magnetic field or of the accelerating gradient, in cw mode. The temperature of the outer cavity surface was constant at 4.3~K for the tests in LHe, whereas it increased with increasing rf field in the test with cryocooler. The inset shows the average cavity temperature as a function of the dissipated power, compared with the temperature of a Cu block with a heater mounted to the 4~K stage of the cryocooler as a function of the heater power.}
\end{figure}

\begin{figure}[htb]
\includegraphics*[width=85mm]{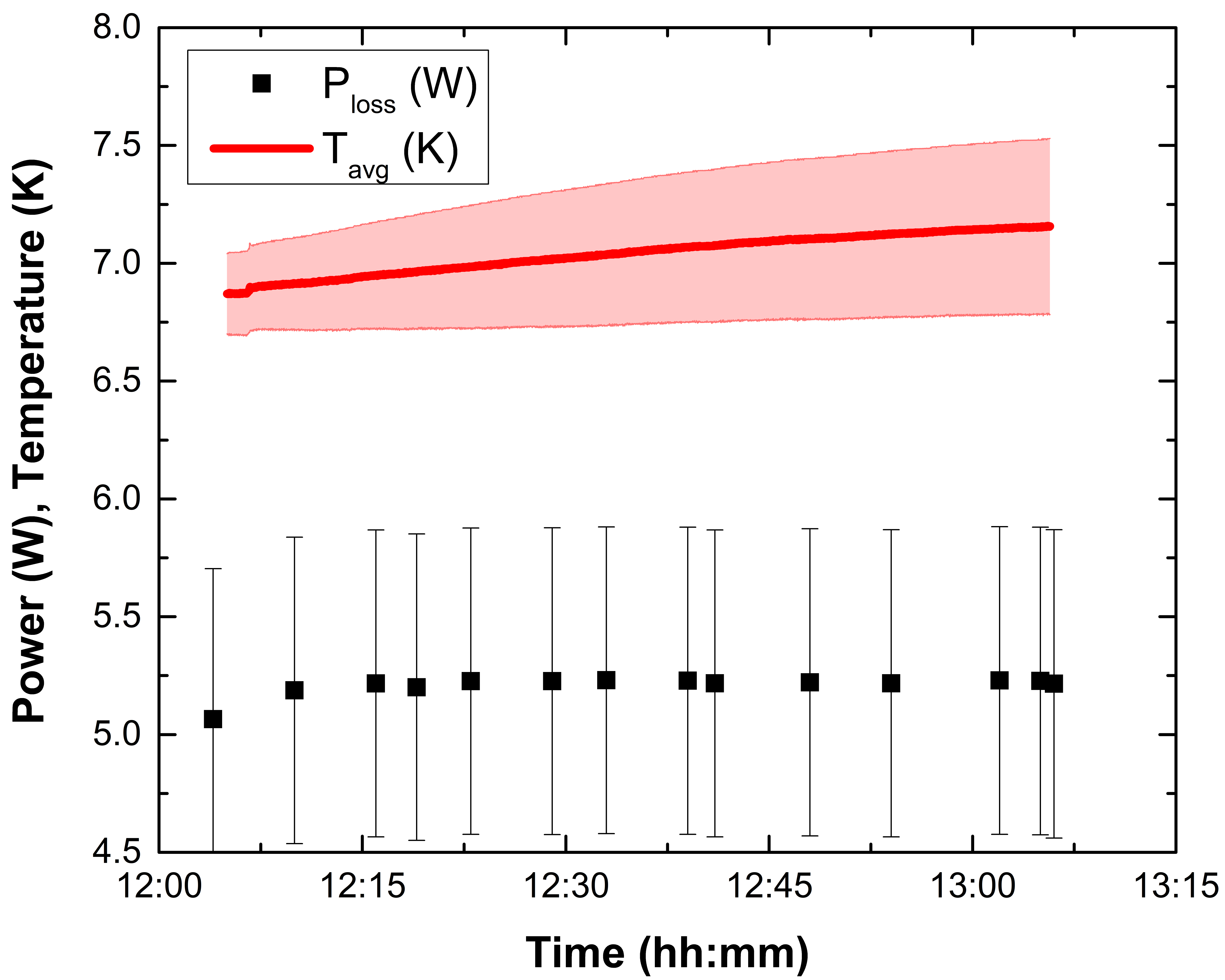}
\caption{\label{fig:1h} Average temperature of the cavity outer surface while operating the conduction cooled cavity in cw mode with a constant dissipated power of 5~W for 1 h with no indication of thermal instability. The width of the shaded area corresponds to the standard deviation.}
\end{figure}

The test stand was inserted in a VTA vertical cryostat, which was used only as a vacuum vessel for this test. The magnitude of the ambient magnetic flux density at the cavity was less than 3~mG. 
%The cooldown rate was $\sim 18$~K/h between 300~K and 70~K, it accelerated to $\sim 1.2$~K/min between 50~K and 10~K and it slowed to $\sim 0.04$~K/h below $\sim 5$~K. 
The cooldown lasted about three days and the average steady state temperature of sensors along the Cu plate attached to the cavity equator ring and on top and bottom of the cavity was $3.8 \pm 0.4$~K. 
%During the cooldown, the magnetic flux measured by one of the FGM probes increased up to 27~mG just before the cavity made the transition to the superconducting state at $\sim 18$~K at which point a jump in the magnetic flux indicated that a fraction of the local magnetic flux might have been trapped in the superconductor. The $Q_0$-value at $B_p = 2.4$~mT was $7 \times 10^9$ and it reduced to $2 \times 10^9$ as the rf field was increased to $B_p = 28.3$~mT, at which point a Q-switch occurred, reducing both $B_p$ and $Q_0$ to 21.9~mT and $4.3 \times 10^8$, respectively, corresponding to $P_{loss}=5$~W. The cavity was held at this level of dissipated power for 1 h, the average temperature holding approximately constant at $\sim 7$~K and without exhibiting any sign of thermal instability, after which the rf power was turned off.
%The cryocooler was turned off and the cavity was let to warm up to $\sim 200$~K at which point the cryocooler was turned back on.
In order to achieve a good thermalization of the cavity in the vicinity of 18~K, which is required in order to minimize rf losses due to trapped magnetic flux generated by thermoelectric currents \cite{Hall}, the cryocooler was cycled on and off twice close to this temperature. The maximum magnetic flux density measured by the FGM probes close to $T_c$ was $\sim 14$~mG and the temperature gradient along the cavity was $\sim 0.09$~K/cm. The $Q_0$-value at $B_p = 2.4$~mT was $1 \times 10^{10}$ and the $Q_0(B_p)$ curve is shown in Fig.~\ref{fig:rf} along with the data measured in LHe at 4.3~K before and after Cu coating. The cavity reached a maximum $B_p$-value of 29~mT above which a Q-switch occur, reducing both $B_p$ and $Q_0$ to 22~mT and $5 \times 10^8$, respectively. It was verified that the $Q_0$ vs. $B_p$ curve is reversible when lowering the forward power. The test was stopped at $P_{loss} = 5$~W,  limited by the power handling capability of the input power cable. 
The cavity was held at this level of dissipated power, corresponding to $Q_0 = 5\times10^8$ at 22~mT, for 1 h after which the rf power was turned off. The average cavity temperature, $T_{avg}$, showed a modest increase from 6.9~K to 7.1~K , as shown in Fig.~\ref{fig:1h}. There was no indication of thermal instability, such as sudden temperature jumps or $dT/dt$ increasing over time during this extended cavity operation test.
All rf tests were done in continuous-wave (cw) mode (100\% duty factor) and there were no detectable X-rays from field-emitted electrons in any of the tests.
The amplitude of the cavity microphonics was measured using the digital low-level rf control system used for the cavity rf test \cite{LLRF} and the peak-to-peak value was 13.8~Hz. The frequency of the microphonics was 1.2~Hz, which is the frequency of the displacer in the 4~K stage of the cryocooler.

\begin{figure}[htb]
\includegraphics*[width=80mm]{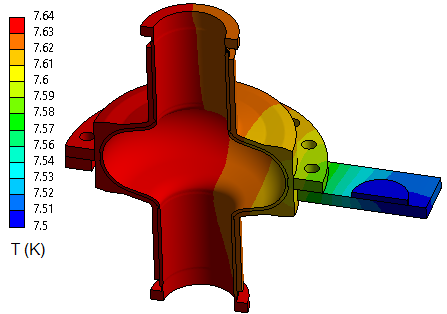}
\caption{\label{fig:sim} Temperature distribution on the cavity surface and Cu plate with 5~W rf heat load and 0.58 W static heat load, calculated with ANSYS.}
\end{figure}

A thermal breakdown of the superconducting state is a common limitation in the operation of SRF cavities in particle accelerators, even when cooled in superfluid He. Except for few exceptions, such thermal breakdown occurs well below the superheating field of the superconductor and it is caused by the presence of defects on the inner cavity surface \cite{Padamsee2}. An example of such defects are normal conducting inclusions which are heated by the rf field and when the local temperature exceeds $T_c$ the surrounding superconductor quenches. Such quenches can occur even at a relatively low power density.
The Q-switch which limits the maximum achievable surface field in this cavity is attributed to defective regions in poor thermal contact with the surrounding superconductor. At the onset of the Q-switch, these regions may become normal conducting and dissipate more and more of the cavity's stored energy as more power is transmitted into the cavity. One possibility, is the presence of a large number of $\mu$m-size defects distributed uniformely over the cavity surface, given the uniformity of the temperature distribution on the cavity outer surface even with a dissipated power of 5~W. Such case was evaluated with a steady-state thermal analysis with ANSYS and the temperature distribution is shown in Fig.~\ref{fig:sim}. A power dissipation of 5~W was applied at the inner surface and an estimated static heat leak of 0.58~W was applied to the bottom of the Cu plate. The contact thermal conductance was set to  0.7~W/K and the temperature at the cryocooler location was set to 7.5~K, based on the heat capacity map.   Figure~\ref{fig:sim} indeed shows that the temperature is quite uniform over the whole cavity surface and close to that of the cryocooler, because of the high-conductivity Cu layer. 

The change in the $Q_0(B_p)$ curve above $\sim20$~mT indicates an onset of ohmic-type losses as $P_{loss} \propto H_p^2$ at higher rf field. If such dependence would have continued, without the occurrence of a Q-switch, a $P_{loss} = 5$~W would have been reached at $B_p = 41$~mT. Such value of $B_p$ would have met the requirement for a single-cell cavity designed for a 1~MeV electron linear accelerator for environmental remediation~\cite{EnvAcc}.
The origin, size and location of the defective regions are unclear at this stage. One possibility is the contamination of the Nb$_3$Sn film by the plating solution. Another possibility is related to strain of the Nb$_3$Sn film given by the differential thermal expansion coefficient between the Cu layer and the Nb layer, since it is well known that the superconducting properties of Nb$_3$Sn are very sensitive to strain \cite{Godeke}. A finite element mechanical analysis with ANSYS showed that stresses as high as $\sim 275$~MPa at the irises and $\sim185$~MPa elsewhere on the cell may be applied to the Nb$_3$Sn-coated Nb, due to the larger thermal contraction of Cu. This aspect and possible mitigations will require further studies.

In summary, we were able to operate a multi-metallic SRF cavity with conduction cooling, using a commercial GM cryocooler, in cw mode up to 29~mT peak surface magnetic field and up to 5~W of power dissipation. In spite of the rigid connection between the cryocooler's 4~K stage and the cavity, the amplitude of the microphonics does not represent an issue, as it was well within what is typically achieved and controlled in SRF accelerator cavities \cite{Powers, Cornell}, particularly considering the low loaded-Q values ($10^4 - 10^5$) typically required for a low-energy, high-power accelerator. To the author's knwoledge, the maximum $B_p$ and $P_{loss}$ values we have reported are the highest ever achieved by a conduction cooled SRF cavity and represent a fundamental stepping stone towards the demonstration of compact, low-cost accelerators for applications in industry, medicine or for university-scale research.

After this work was completed, we became aware of a new report from Fermilab in which $B_p \sim 24$~mT and a $P_{loss}$ of 4~W were achieved in a cryocooler conduction cooled 650~MHz single-cell cavity \cite{Dhuley}

\section*{Acknowledgments}
The authors woud like to acknowledge B. Golesich at CTC, A. Tuck at AJ Tuck Co. and D. Combs of JLab's machine shop for their expert craftsmanship to produce the cavity's thick Cu shell. We would like to thank our colleagues K. Harding and J. Henry for the design of the cryocooler test stand, E. Daly for helpful discussions on the test setup, JLab's Cavity Production Group for helping with cleaning of the parts and the cavity and D. Tucker for helping with the assembly of the test stand. We would like to thank I. Parajuli of Old Dominion University for helping with the cavity assembly and sensors installation. We would also like to acknowledge Dr. G. Eremeev of Fermilab, formerly at JLab, for allowing access to the Nb$_3$Sn coating furnace and M. Dale of Sumitomo Cryogenics of America for lending the cryocooler used for this study.
This work is authored by Jefferson Science Associates, LLC under U.S. DOE Contract No. DE-AC05-06OR23177 and it is partly supported by the Presidential Early Career Award of G. Ciovati. The work of U. Pudasaini was supported by the DOE Early Career Award of G. Eremeev.

\bibliography{PRApp_Cond_cool_Bib}

\begin{thebibliography}{26}%
\makeatletter
\providecommand \@ifxundefined [1]{%
 \@ifx{#1\undefined}
}%
\providecommand \@ifnum [1]{%
 \ifnum #1\expandafter \@firstoftwo
 \else \expandafter \@secondoftwo
 \fi
}%
\providecommand \@ifx [1]{%
 \ifx #1\expandafter \@firstoftwo
 \else \expandafter \@secondoftwo
 \fi
}%
\providecommand \natexlab [1]{#1}%
\providecommand \enquote  [1]{``#1''}%
\providecommand \bibnamefont  [1]{#1}%
\providecommand \bibfnamefont [1]{#1}%
\providecommand \citenamefont [1]{#1}%
\providecommand \href@noop [0]{\@secondoftwo}%
\providecommand \href [0]{\begingroup \@sanitize@url \@href}%
\providecommand \@href[1]{\@@startlink{#1}\@@href}%
\providecommand \@@href[1]{\endgroup#1\@@endlink}%
\providecommand \@sanitize@url [0]{\catcode `\\12\catcode `\$12\catcode
  `\&12\catcode `\#12\catcode `\^12\catcode `\_12\catcode `\%12\relax}%
\providecommand \@@startlink[1]{}%
\providecommand \@@endlink[0]{}%
\providecommand \url  [0]{\begingroup\@sanitize@url \@url }%
\providecommand \@url [1]{\endgroup\@href {#1}{\urlprefix }}%
\providecommand \urlprefix  [0]{URL }%
\providecommand \Eprint [0]{\href }%
\providecommand \doibase [0]{http://dx.doi.org/}%
\providecommand \selectlanguage [0]{\@gobble}%
\providecommand \bibinfo  [0]{\@secondoftwo}%
\providecommand \bibfield  [0]{\@secondoftwo}%
\providecommand \translation [1]{[#1]}%
\providecommand \BibitemOpen [0]{}%
\providecommand \bibitemStop [0]{}%
\providecommand \bibitemNoStop [0]{.\EOS\space}%
\providecommand \EOS [0]{\spacefactor3000\relax}%
\providecommand \BibitemShut  [1]{\csname bibitem#1\endcsname}%
\let\auto@bib@innerbib\@empty
%</preamble>
\bibitem [{\citenamefont {Padamsee}(2009{\natexlab{a}})}]{Padamsee}%
  \BibitemOpen
  \bibfield  {author} {\bibinfo {author} {\bibfnamefont {H.}~\bibnamefont
  {Padamsee}},\ }\enquote {\bibinfo {title} {Applications and operations},}\
  in\ \href {\doibase 10.1002/9783527627172.ch11} {\emph {\bibinfo {booktitle}
  {RF Superconductivity}}}\ (\bibinfo  {publisher} {John Wiley \& Sons, Ltd},\
  \bibinfo {year} {2009})\ Chap.~\bibinfo {chapter} {11}, pp.\ \bibinfo {pages}
  {333--387}\BibitemShut {NoStop}%
\bibitem [{\citenamefont {Belomestnykh}(2013)}]{RAST_hibeta}%
  \BibitemOpen
  \bibfield  {author} {\bibinfo {author} {\bibfnamefont {S.}~\bibnamefont
  {Belomestnykh}},\ }in\ \href {\doibase 10.1142/S1793626812300006X} {\emph
  {\bibinfo {booktitle} {Reviews of Accelerator Science and Technology}}},\
  Vol.~\bibinfo {volume} {5},\ \bibinfo {editor} {edited by\ \bibinfo {editor}
  {\bibfnamefont {A.~W.}\ \bibnamefont {Chao}}\ and\ \bibinfo {editor}
  {\bibfnamefont {W.}~\bibnamefont {Chou}}}\ (\bibinfo  {publisher} {WORLD
  SCIENTIFIC},\ \bibinfo {address} {Singapore},\ \bibinfo {year} {2013})\ pp.\
  \bibinfo {pages} {147--184}\BibitemShut {NoStop}%
\bibitem [{\citenamefont {Kelly}(2013)}]{RAST_lowbeta}%
  \BibitemOpen
  \bibfield  {author} {\bibinfo {author} {\bibfnamefont {M.}~\bibnamefont
  {Kelly}},\ }in\ \href {\doibase 10.1142/S1793626812300071} {\emph {\bibinfo
  {booktitle} {Reviews of Accelerator Science and Technology}}},\ Vol.~\bibinfo
  {volume} {5},\ \bibinfo {editor} {edited by\ \bibinfo {editor} {\bibfnamefont
  {A.~W.}\ \bibnamefont {Chao}}\ and\ \bibinfo {editor} {\bibfnamefont
  {W.}~\bibnamefont {Chou}}}\ (\bibinfo  {publisher} {WORLD SCIENTIFIC},\
  \bibinfo {address} {Singapore},\ \bibinfo {year} {2013})\ pp.\ \bibinfo
  {pages} {185--203}\BibitemShut {NoStop}%
\bibitem [{\citenamefont {Boulware}()}]{Niowave}%
  \BibitemOpen
  \bibfield  {author} {\bibinfo {author} {\bibfnamefont {C.}~\bibnamefont
  {Boulware}},\ }\href@noop {} {\enquote {\bibinfo {title} {4 {K}
  superconducting linacs for commercial applications},}\ }\bibinfo
  {howpublished} {presented at 2016 North American Particle Accelerator
  Conference, Chicago, IL, October 2016}\BibitemShut {NoStop}%
\bibitem [{\citenamefont {Minehara}\ \emph {et~al.}()\citenamefont {Minehara},
  \citenamefont {Kikuzawa}, \citenamefont {Nagai}, \citenamefont {Kato},
  \citenamefont {Sawamura}, \citenamefont {Takao}, \citenamefont {Sugimoto},
  \citenamefont {Ohkubo}, \citenamefont {Sasabe}, \citenamefont {Suzuki},\ and\
  \citenamefont {Kawarasaki}}]{JAERI1}%
  \BibitemOpen
  \bibfield  {author} {\bibinfo {author} {\bibfnamefont {E.~J.}\ \bibnamefont
  {Minehara}}, \bibinfo {author} {\bibfnamefont {N.}~\bibnamefont {Kikuzawa}},
  \bibinfo {author} {\bibfnamefont {R.}~\bibnamefont {Nagai}}, \bibinfo
  {author} {\bibfnamefont {R.}~\bibnamefont {Kato}}, \bibinfo {author}
  {\bibfnamefont {M.}~\bibnamefont {Sawamura}}, \bibinfo {author}
  {\bibfnamefont {M.}~\bibnamefont {Takao}}, \bibinfo {author} {\bibfnamefont
  {M.}~\bibnamefont {Sugimoto}}, \bibinfo {author} {\bibfnamefont
  {M.}~\bibnamefont {Ohkubo}}, \bibinfo {author} {\bibfnamefont
  {J.}~\bibnamefont {Sasabe}}, \bibinfo {author} {\bibfnamefont
  {Y.}~\bibnamefont {Suzuki}}, \ and\ \bibinfo {author} {\bibfnamefont
  {Y.}~\bibnamefont {Kawarasaki}},\ }in\ \href@noop {} {\emph {\bibinfo
  {booktitle} {{Proc. 6th Int. Conf. RF Superconductivity (SRF’93), Newport
  News, VA, USA, Oct. 1993, paper SRF93I34}}}}\ (\bibinfo  {publisher} {JACoW
  Publishing, Geneva, Switzerland})\ pp.\ \bibinfo {pages}
  {886--894}\BibitemShut {NoStop}%
\bibitem [{\citenamefont {Kikuzawa}\ \emph {et~al.}()\citenamefont {Kikuzawa},
  \citenamefont {Nagai}, \citenamefont {Sawamura}, \citenamefont {Nishimori},
  \citenamefont {Minehara},\ and\ \citenamefont {Suzuki}}]{JAERI2}%
  \BibitemOpen
  \bibfield  {author} {\bibinfo {author} {\bibfnamefont {N.}~\bibnamefont
  {Kikuzawa}}, \bibinfo {author} {\bibfnamefont {R.}~\bibnamefont {Nagai}},
  \bibinfo {author} {\bibfnamefont {M.}~\bibnamefont {Sawamura}}, \bibinfo
  {author} {\bibfnamefont {N.}~\bibnamefont {Nishimori}}, \bibinfo {author}
  {\bibfnamefont {E.}~\bibnamefont {Minehara}}, \ and\ \bibinfo {author}
  {\bibfnamefont {Y.}~\bibnamefont {Suzuki}},\ }in\ \href@noop {} {\emph
  {\bibinfo {booktitle} {{Proc. 8th Int. Conf. RF Superconductivity (SRF’97),
  Padova, Italy, Oct. 1997, paper SRF97C40}}}}\ (\bibinfo  {publisher} {JACoW
  Publishing, Geneva, Switzerland})\ pp.\ \bibinfo {pages}
  {769--773}\BibitemShut {NoStop}%
\bibitem [{\citenamefont {Posen}\ and\ \citenamefont
  {Hall}(2017)}]{Posen_2017}%
  \BibitemOpen
  \bibfield  {author} {\bibinfo {author} {\bibfnamefont {S.}~\bibnamefont
  {Posen}}\ and\ \bibinfo {author} {\bibfnamefont {D.~L.}\ \bibnamefont
  {Hall}},\ }\href {\doibase 10.1088/1361-6668/30/3/033004} {\bibfield
  {journal} {\bibinfo  {journal} {Superconductor Science and Technology}\
  }\textbf {\bibinfo {volume} {30}},\ \bibinfo {pages} {033004} (\bibinfo
  {year} {2017})}\BibitemShut {NoStop}%
\bibitem [{\citenamefont {Kephart}\ \emph {et~al.}()\citenamefont {Kephart},
  \citenamefont {Chase}, \citenamefont {Gonin}, \citenamefont {Grassellino},
  \citenamefont {Kazakov}, \citenamefont {Khabiboulline}, \citenamefont
  {Nagaitsev}, \citenamefont {Pasquinelli}, \citenamefont {Posen},
  \citenamefont {Pronitchev}, \citenamefont {Romanenko}, \citenamefont
  {Yakovlev}, \citenamefont {Biedron}, \citenamefont {Milton}, \citenamefont
  {Sipahi}, \citenamefont {Chattopadhyay},\ and\ \citenamefont
  {Piot}}]{Kephart}%
  \BibitemOpen
  \bibfield  {author} {\bibinfo {author} {\bibfnamefont {R.}~\bibnamefont
  {Kephart}}, \bibinfo {author} {\bibfnamefont {B.}~\bibnamefont {Chase}},
  \bibinfo {author} {\bibfnamefont {I.}~\bibnamefont {Gonin}}, \bibinfo
  {author} {\bibfnamefont {A.}~\bibnamefont {Grassellino}}, \bibinfo {author}
  {\bibfnamefont {S.}~\bibnamefont {Kazakov}}, \bibinfo {author} {\bibfnamefont
  {T.}~\bibnamefont {Khabiboulline}}, \bibinfo {author} {\bibfnamefont
  {S.}~\bibnamefont {Nagaitsev}}, \bibinfo {author} {\bibfnamefont
  {R.}~\bibnamefont {Pasquinelli}}, \bibinfo {author} {\bibfnamefont
  {S.}~\bibnamefont {Posen}}, \bibinfo {author} {\bibfnamefont
  {O.}~\bibnamefont {Pronitchev}}, \bibinfo {author} {\bibfnamefont
  {A.}~\bibnamefont {Romanenko}}, \bibinfo {author} {\bibfnamefont
  {V.}~\bibnamefont {Yakovlev}}, \bibinfo {author} {\bibfnamefont
  {S.}~\bibnamefont {Biedron}}, \bibinfo {author} {\bibfnamefont
  {S.}~\bibnamefont {Milton}}, \bibinfo {author} {\bibfnamefont
  {N.}~\bibnamefont {Sipahi}}, \bibinfo {author} {\bibfnamefont
  {S.}~\bibnamefont {Chattopadhyay}}, \ and\ \bibinfo {author} {\bibfnamefont
  {P.}~\bibnamefont {Piot}},\ }in\ \href@noop {} {\emph {\bibinfo {booktitle}
  {{Proc. 17th Int. Conf. RF Superconductivity (SRF’15), Whistler, Canada,
  Sep. 2015, paper FRBA03}}}}\ (\bibinfo  {publisher} {JACoW Publishing,
  Geneva, Switzerland})\ pp.\ \bibinfo {pages} {1467--1473}\BibitemShut
  {NoStop}%
\bibitem [{\citenamefont {Barletta}\ and\ \citenamefont
  {Borland}(2010)}]{CLS_Report}%
  \BibitemOpen
  \bibfield  {author} {\bibinfo {author} {\bibfnamefont {W.}~\bibnamefont
  {Barletta}}\ and\ \bibinfo {author} {\bibfnamefont {M.}~\bibnamefont
  {Borland}},\ }\href@noop {} {\emph {\bibinfo {title} {Report of the Basic
  Energy Sciences Workshop on Compact Light Sources, Rockville, MD}}},\
  \bibinfo {type} {Tech. Rep.}\ (\bibinfo  {institution} {{U.S.} {D}epartment
  of {E}nergy},\ \bibinfo {year} {2010})\BibitemShut {NoStop}%
\bibitem [{\citenamefont {Ciovati}\ \emph {et~al.}(2018)\citenamefont
  {Ciovati}, \citenamefont {Anderson}, \citenamefont {Coriton}, \citenamefont
  {Guo}, \citenamefont {Hannon}, \citenamefont {Holland}, \citenamefont
  {LeSher}, \citenamefont {Marhauser}, \citenamefont {Rathke}, \citenamefont
  {Rimmer}, \citenamefont {Schultheiss},\ and\ \citenamefont {Vylet}}]{EnvAcc}%
  \BibitemOpen
  \bibfield  {author} {\bibinfo {author} {\bibfnamefont {G.}~\bibnamefont
  {Ciovati}}, \bibinfo {author} {\bibfnamefont {J.}~\bibnamefont {Anderson}},
  \bibinfo {author} {\bibfnamefont {B.}~\bibnamefont {Coriton}}, \bibinfo
  {author} {\bibfnamefont {J.}~\bibnamefont {Guo}}, \bibinfo {author}
  {\bibfnamefont {F.}~\bibnamefont {Hannon}}, \bibinfo {author} {\bibfnamefont
  {L.}~\bibnamefont {Holland}}, \bibinfo {author} {\bibfnamefont
  {M.}~\bibnamefont {LeSher}}, \bibinfo {author} {\bibfnamefont
  {F.}~\bibnamefont {Marhauser}}, \bibinfo {author} {\bibfnamefont
  {J.}~\bibnamefont {Rathke}}, \bibinfo {author} {\bibfnamefont
  {R.}~\bibnamefont {Rimmer}}, \bibinfo {author} {\bibfnamefont
  {T.}~\bibnamefont {Schultheiss}}, \ and\ \bibinfo {author} {\bibfnamefont
  {V.}~\bibnamefont {Vylet}},\ }\href {\doibase
  10.1103/PhysRevAccelBeams.21.091601} {\bibfield  {journal} {\bibinfo
  {journal} {Phys. Rev. Accel. Beams}\ }\textbf {\bibinfo {volume} {21}},\
  \bibinfo {pages} {091601} (\bibinfo {year} {2018})}\BibitemShut {NoStop}%
\bibitem [{\citenamefont {Dhuley}\ \emph {et~al.}(shed)\citenamefont {Dhuley},
  \citenamefont {Geelhoed}, \citenamefont {Zhao}, \citenamefont {Terechkine},
  \citenamefont {Alvarez}, \citenamefont {Prokofiev},\ and\ \citenamefont
  {Thangaraj}}]{Dhuley1}%
  \BibitemOpen
  \bibfield  {author} {\bibinfo {author} {\bibfnamefont {R.~C.}\ \bibnamefont
  {Dhuley}}, \bibinfo {author} {\bibfnamefont {M.~I.}\ \bibnamefont
  {Geelhoed}}, \bibinfo {author} {\bibfnamefont {Y.}~\bibnamefont {Zhao}},
  \bibinfo {author} {\bibfnamefont {I.}~\bibnamefont {Terechkine}}, \bibinfo
  {author} {\bibfnamefont {M.}~\bibnamefont {Alvarez}}, \bibinfo {author}
  {\bibfnamefont {O.}~\bibnamefont {Prokofiev}}, \ and\ \bibinfo {author}
  {\bibfnamefont {J.~C.~T.}\ \bibnamefont {Thangaraj}},\ }in\ \href
  {https://lss.fnal.gov/archive/2019/conf/fermilab-conf-19-351-di-ldrd-td.pdf}
  {\emph {\bibinfo {booktitle} {{IOP} {C}onference {S}eries: {M}aterials
  {S}cience and {E}ngineering}}}\ (\bibinfo {year} {to be
  published})\BibitemShut {NoStop}%
\bibitem [{\citenamefont {Duhley}()}]{DhuleyPC}%
  \BibitemOpen
  \bibfield  {author} {\bibinfo {author} {\bibfnamefont {R.}~\bibnamefont
  {Duhley}},\ }\href@noop {} {}\bibinfo {howpublished} {personal
  communication}\BibitemShut {NoStop}%
\bibitem [{\citenamefont {Sekutowicz}()}]{HG}%
  \BibitemOpen
  \bibfield  {author} {\bibinfo {author} {\bibfnamefont {J.}~\bibnamefont
  {Sekutowicz}},\ }in\ \href@noop {} {\emph {\bibinfo {booktitle} {{Proc. 20th
  Particle Accelerator Conf. (PAC’03), Portland, OR, USA, May 2003, paper
  TPAB085}}}}\ (\bibinfo  {publisher} {JACoW Publishing, Geneva, Switzerland})\
  pp.\ \bibinfo {pages} {1395--1397}\BibitemShut {NoStop}%
\bibitem [{\citenamefont {Eremeev}\ \emph {et~al.}()\citenamefont {Eremeev},
  \citenamefont {Clemens}, \citenamefont {Macha}, \citenamefont {Park},\ and\
  \citenamefont {Williams}}]{Nb3Sn_furnace}%
  \BibitemOpen
  \bibfield  {author} {\bibinfo {author} {\bibfnamefont {G.~V.}\ \bibnamefont
  {Eremeev}}, \bibinfo {author} {\bibfnamefont {W.~A.}\ \bibnamefont
  {Clemens}}, \bibinfo {author} {\bibfnamefont {K.}~\bibnamefont {Macha}},
  \bibinfo {author} {\bibfnamefont {H.}~\bibnamefont {Park}}, \ and\ \bibinfo
  {author} {\bibfnamefont {R.~S.}\ \bibnamefont {Williams}},\ }in\ \href
  {\doibase 10.18429/JACoW-IPAC2015-WEPWI011} {\emph {\bibinfo {booktitle}
  {{Proc. 6th Int. Particle Accelerator Conf. (IPAC’15), Richmond, VA, USA,
  May 2015}}}}\ (\bibinfo  {publisher} {JACoW Publishing, Geneva,
  Switzerland})\ pp.\ \bibinfo {pages} {3512--3514}\BibitemShut {NoStop}%
\bibitem [{\citenamefont {Benvenuti}\ \emph {et~al.}()\citenamefont
  {Benvenuti}, \citenamefont {Bernard}, \citenamefont {Bloess}, \citenamefont
  {Cavallari}, \citenamefont {Chiaveri}, \citenamefont {Haebel}, \citenamefont
  {Hilleret}, \citenamefont {Tückmantel},\ and\ \citenamefont
  {Weingarten}}]{LEP}%
  \BibitemOpen
  \bibfield  {author} {\bibinfo {author} {\bibfnamefont {C.}~\bibnamefont
  {Benvenuti}}, \bibinfo {author} {\bibfnamefont {P.}~\bibnamefont {Bernard}},
  \bibinfo {author} {\bibfnamefont {D.}~\bibnamefont {Bloess}}, \bibinfo
  {author} {\bibfnamefont {G.}~\bibnamefont {Cavallari}}, \bibinfo {author}
  {\bibfnamefont {E.}~\bibnamefont {Chiaveri}}, \bibinfo {author}
  {\bibfnamefont {E.}~\bibnamefont {Haebel}}, \bibinfo {author} {\bibfnamefont
  {N.}~\bibnamefont {Hilleret}}, \bibinfo {author} {\bibfnamefont
  {J.}~\bibnamefont {Tückmantel}}, \ and\ \bibinfo {author} {\bibfnamefont
  {W.}~\bibnamefont {Weingarten}},\ }in\ \href {\doibase
  10.1109/PAC.1991.164525} {\emph {\bibinfo {booktitle} {{Proc. 14th Particle
  Accelerator Conf. (PAC’91), San Francisco, CA, USA, May 1991}}}}\ (\bibinfo
   {publisher} {JACoW Publishing, Geneva, Switzerland})\ pp.\ \bibinfo {pages}
  {1023--1026}\BibitemShut {NoStop}%
\bibitem [{\citenamefont {Ciovati}\ \emph {et~al.}()\citenamefont {Ciovati},
  \citenamefont {Cheng}, \citenamefont {Daly}, \citenamefont {Eremeev},
  \citenamefont {Henry}, , \citenamefont {Pudasaini}, \citenamefont
  {Parajuli},\ and\ \citenamefont {Rimmer}}]{Gigi_SRF19}%
  \BibitemOpen
  \bibfield  {author} {\bibinfo {author} {\bibfnamefont {G.}~\bibnamefont
  {Ciovati}}, \bibinfo {author} {\bibfnamefont {G.}~\bibnamefont {Cheng}},
  \bibinfo {author} {\bibfnamefont {E.}~\bibnamefont {Daly}}, \bibinfo {author}
  {\bibfnamefont {G.}~\bibnamefont {Eremeev}}, \bibinfo {author} {\bibfnamefont
  {J.}~\bibnamefont {Henry}}, , \bibinfo {author} {\bibfnamefont
  {U.}~\bibnamefont {Pudasaini}}, \bibinfo {author} {\bibfnamefont
  {I.}~\bibnamefont {Parajuli}}, \ and\ \bibinfo {author} {\bibfnamefont
  {R.}~\bibnamefont {Rimmer}},\ }in\ \href {\doibase
  10.18429/JACoW-SRF2019-TUP050} {\emph {\bibinfo {booktitle} {{Proc. 19th Int.
  Conf. RF Superconductivity (SRF’19), Dresden, Germany, Jun.-Jul. 2019,
  paper TUP050}}}}\ (\bibinfo  {publisher} {JACoW Publishing, Geneva,
  Switzerland})\ pp.\ \bibinfo {pages} {540--544}\BibitemShut {NoStop}%
\bibitem [{\citenamefont {Kanareykin}()}]{Euclid}%
  \BibitemOpen
  \bibfield  {author} {\bibinfo {author} {\bibfnamefont {A.}~\bibnamefont
  {Kanareykin}},\ }\href@noop {} {}\bibinfo {howpublished} {personal
  communication}\BibitemShut {NoStop}%
\bibitem [{\citenamefont {ANSYS}(2019)}]{ansys}%
  \BibitemOpen
  \bibfield  {author} {\bibinfo {author} {\bibnamefont {ANSYS}},\ }\href
  {ansys.com} {\enquote {\bibinfo {title} {3{D} engineering and designing
  software},}\ } (\bibinfo {year} {2019})\BibitemShut {NoStop}%
\bibitem [{\citenamefont {Dillon}\ \emph {et~al.}(2017)\citenamefont {Dillon},
  \citenamefont {McCusker}, \citenamefont {Dyke}, \citenamefont {Isler},\ and\
  \citenamefont {Christiansen}}]{Dillon_2017}%
  \BibitemOpen
  \bibfield  {author} {\bibinfo {author} {\bibfnamefont {A.}~\bibnamefont
  {Dillon}}, \bibinfo {author} {\bibfnamefont {K.}~\bibnamefont {McCusker}},
  \bibinfo {author} {\bibfnamefont {J.~V.}\ \bibnamefont {Dyke}}, \bibinfo
  {author} {\bibfnamefont {B.}~\bibnamefont {Isler}}, \ and\ \bibinfo {author}
  {\bibfnamefont {M.}~\bibnamefont {Christiansen}},\ }\href {\doibase
  10.1088/1757-899x/278/1/012054} {\bibfield  {journal} {\bibinfo  {journal}
  {{IOP} Conference Series: Materials Science and Engineering}\ }\textbf
  {\bibinfo {volume} {278}},\ \bibinfo {pages} {012054} (\bibinfo {year}
  {2017})}\BibitemShut {NoStop}%
\bibitem [{\citenamefont {Hall}\ \emph {et~al.}()\citenamefont {Hall},
  \citenamefont {Liepe}, \citenamefont {Liarte},\ and\ \citenamefont
  {Sethna}}]{Hall}%
  \BibitemOpen
  \bibfield  {author} {\bibinfo {author} {\bibfnamefont {D.}~\bibnamefont
  {Hall}}, \bibinfo {author} {\bibfnamefont {M.}~\bibnamefont {Liepe}},
  \bibinfo {author} {\bibfnamefont {D.}~\bibnamefont {Liarte}}, \ and\ \bibinfo
  {author} {\bibfnamefont {J.~P.}\ \bibnamefont {Sethna}},\ }in\ \href
  {\doibase 10.18429/JACoW-IPAC2017-MOPVA118} {\emph {\bibinfo {booktitle}
  {{Proc. 8th Int. Particle Accelerator Conf. (IPAC’17), Copenhagen, Denmark,
  May 2017}}}}\ (\bibinfo  {publisher} {JACoW Publishing, Geneva,
  Switzerland})\ pp.\ \bibinfo {pages} {1127--1129}\BibitemShut {NoStop}%
\bibitem [{\citenamefont {Hovater}\ \emph {et~al.}()\citenamefont {Hovater},
  \citenamefont {Allison}, \citenamefont {Bachimanchi}, \citenamefont {Lahti},
  \citenamefont {Musson}, \citenamefont {Plawski}, \citenamefont {Seaton},\
  and\ \citenamefont {Seidman}}]{LLRF}%
  \BibitemOpen
  \bibfield  {author} {\bibinfo {author} {\bibfnamefont {C.}~\bibnamefont
  {Hovater}}, \bibinfo {author} {\bibfnamefont {T.}~\bibnamefont {Allison}},
  \bibinfo {author} {\bibfnamefont {R.}~\bibnamefont {Bachimanchi}}, \bibinfo
  {author} {\bibfnamefont {G.}~\bibnamefont {Lahti}}, \bibinfo {author}
  {\bibfnamefont {J.}~\bibnamefont {Musson}}, \bibinfo {author} {\bibfnamefont
  {T.~E.}\ \bibnamefont {Plawski}}, \bibinfo {author} {\bibfnamefont
  {C.}~\bibnamefont {Seaton}}, \ and\ \bibinfo {author} {\bibfnamefont
  {D.}~\bibnamefont {Seidman}},\ }in\ \href@noop {} {\emph {\bibinfo
  {booktitle} {{Proc. 25th Linear Accelerator Conf. (LINAC’10), Tsukuba,
  Japan, Sep. 2010, paper MOP095}}}}\ (\bibinfo  {publisher} {JACoW Publishing,
  Geneva, Switzerland})\ pp.\ \bibinfo {pages} {280--282}\BibitemShut {NoStop}%
\bibitem [{\citenamefont {Padamsee}(2009{\natexlab{b}})}]{Padamsee2}%
  \BibitemOpen
  \bibfield  {author} {\bibinfo {author} {\bibfnamefont {H.}~\bibnamefont
  {Padamsee}},\ }\enquote {\bibinfo {title} {High-field {Q}-slope and quench
  field},}\ in\ \href {\doibase 10.1002/9783527627172.ch11} {\emph {\bibinfo
  {booktitle} {RF Superconductivity}}}\ (\bibinfo  {publisher} {John Wiley \&
  Sons, Ltd},\ \bibinfo {year} {2009})\ Chap.~\bibinfo {chapter} {5}, pp.\
  \bibinfo {pages} {192--200}\BibitemShut {NoStop}%
\bibitem [{\citenamefont {Godeke}\ \emph {et~al.}(2018)\citenamefont {Godeke},
  \citenamefont {Hellman}, \citenamefont {ten Kate},\ and\ \citenamefont
  {Mentink}}]{Godeke}%
  \BibitemOpen
  \bibfield  {author} {\bibinfo {author} {\bibfnamefont {A.}~\bibnamefont
  {Godeke}}, \bibinfo {author} {\bibfnamefont {F.}~\bibnamefont {Hellman}},
  \bibinfo {author} {\bibfnamefont {H.~H.~J.}\ \bibnamefont {ten Kate}}, \ and\
  \bibinfo {author} {\bibfnamefont {M.~G.~T.}\ \bibnamefont {Mentink}},\ }\href
  {\doibase 10.1088/1361-6668/aad980} {\bibfield  {journal} {\bibinfo
  {journal} {Superconductor Science and Technology}\ }\textbf {\bibinfo
  {volume} {31}},\ \bibinfo {pages} {105011} (\bibinfo {year}
  {2018})}\BibitemShut {NoStop}%
\bibitem [{\citenamefont {Powers}\ \emph {et~al.}(2019)\citenamefont {Powers},
  \citenamefont {Brock},\ and\ \citenamefont {Davis}}]{Powers}%
  \BibitemOpen
  \bibfield  {author} {\bibinfo {author} {\bibfnamefont {T.}~\bibnamefont
  {Powers}}, \bibinfo {author} {\bibfnamefont {N.}~\bibnamefont {Brock}}, \
  and\ \bibinfo {author} {\bibfnamefont {G.}~\bibnamefont {Davis}},\ }in\ \href
  {\doibase doi:10.18429/JACoW-SRF2019-TUP034} {\emph {\bibinfo {booktitle}
  {Proc. SRF'19}}},\ \bibinfo {series and number} {\bibinfo {series}
  {International Conference on RF Superconductivity}\ No.~\bibinfo {number}
  {19}}\ (\bibinfo  {publisher} {JACoW Publishing, Geneva, Switzerland},\
  \bibinfo {year} {2019})\ pp.\ \bibinfo {pages} {493--498},\ \bibinfo {note}
  {https://doi.org/10.18429/JACoW-SRF2019-TUP034}\BibitemShut {NoStop}%
\bibitem [{\citenamefont {Liepe}\ and\ \citenamefont
  {Belomestnykh}()}]{Cornell}%
  \BibitemOpen
  \bibfield  {author} {\bibinfo {author} {\bibfnamefont {M.}~\bibnamefont
  {Liepe}}\ and\ \bibinfo {author} {\bibfnamefont {S.}~\bibnamefont
  {Belomestnykh}},\ }in\ \href@noop {} {\emph {\bibinfo {booktitle} {{Proc.
  20th Particle Accelerator Conf. (PAC’03), Portland, OR, USA, May 2003,
  paper TPAB055}}}}\ (\bibinfo  {publisher} {JACoW Publishing, Geneva,
  Switzerland})\ pp.\ \bibinfo {pages} {1326--1328}\BibitemShut {NoStop}%
\bibitem [{\citenamefont {Dhuley}\ \emph {et~al.}()\citenamefont {Dhuley},
  \citenamefont {Posen}, \citenamefont {Geelhoed}, \citenamefont {Prokofiev},\
  and\ \citenamefont {Thangaraj}}]{Dhuley}%
  \BibitemOpen
  \bibfield  {author} {\bibinfo {author} {\bibfnamefont {R.~C.}\ \bibnamefont
  {Dhuley}}, \bibinfo {author} {\bibfnamefont {S.}~\bibnamefont {Posen}},
  \bibinfo {author} {\bibfnamefont {M.~I.}\ \bibnamefont {Geelhoed}}, \bibinfo
  {author} {\bibfnamefont {O.}~\bibnamefont {Prokofiev}}, \ and\ \bibinfo
  {author} {\bibfnamefont {J.~C.~T.}\ \bibnamefont {Thangaraj}},\ }\href@noop
  {} {\enquote {\bibinfo {title} {Demonstration of a cryocooler
  conduction-cooled superconducting radiofrequency cavity operating at
  practical cw accelerating gradients},}\ }\Eprint
  {http://arxiv.org/abs/2001.0782} {arXiv:2001.0782 [physics.ins-det]}
  \BibitemShut {NoStop}%
\end{thebibliography}%

\end{document}